# Doppler-shifted X-ray Spectroscopy of Nonradiative Electron Capture in Relativistic Collisions of $Xe^{54+}$ Ions with Kr and Xe Atoms

Bian Yang,[1, 2] Deyang Yu,[1, 2, *] Konstantin N. Lyashchenko,[3, 4] Caojie Shao,[1, 2] Zhongwen Wu,[5] Mingwu Zhang,[1, 2] Oleg Yu. Andreev,[3] Junliang Liu,[1, 2] Zhangyong Song,[1, 2] Yingli Xue,[1, 2] Wei Wang,[1, 2] Fangfang Ruan,[6] Yehong Wu,[7] Rongchun Lu,[1] Chenzhong Dong,[5] and Xiaohong Cai[1]

[1] *Institute of Modern Physics, Chinese Academy of Sciences, Lanzhou 730000, China*

[2] *University of Chinese Academy of Sciences, Beijing 100049, China*

[3] *Department of Physics, St. Petersburg State University, 7/9 Universitetskaya nab., St. Petersburg, 199034, Russia*

[4] *Petersburg Nuclear Physics Institute named by B.P. Konstantinov of National Research Centre "Kurchatov Institute," mkr. Orlova roshcha 1, Gatchina, 188300, Leningrad District, Russia*

[5] *Key Laboratory of Atomic and Molecular Physics & Functional Materials of Gansu Province, College of Physics and Electronic Engineering, Northwest Normal University, Lanzhou 730070, China*

[6] *Department of Medical Imaging, Hangzhou Medical College, Hangzhou 310053, China*

[7] *School of Basic Medical Sciences, Shanxi Medical University, Taiyuan 030001, China*

**Abstract:** We present an angular-resolved Doppler spectroscopy study of nonradiative electron capture in relativistic collisions of bare $Xe^{54+}$ ions with Kr and Xe gas targets at the HIRFL-CSR storage ring. The energy spectra and angular distributions of X-rays emitted from fast-moving down-charged projectiles were measured at five observation angles of 35°, 60°, 90°, 120°, and 145° and three collision energies of 95, 146, and 197 MeV/u by employing the effect of Doppler shift. The transition intensities of $Xe^{53+}$ ions with small energy differences were precisely determined. In symmetric $Xe^{54+}$→Xe collisions, the transition intensities of $Xe^{53+}$ and $Xe^{52+}$ ions were identified when X-rays emitted by projectiles overlapped with *K* X-rays arising from target ionization. The anisotropy parameters of the $K\alpha_1$(+M2) transition were derived from the angular emission patterns of the corresponding spectral lines. The relative populations of the *L*, *M*, and *N*-shell excited levels of $Xe^{53+}$ and $Xe^{52+}$ were further deduced from the intensity ratios of I(Ly-β)/I(Ly-α), I(Ly-γ)/I(Ly-α), and I(Kα)/I(Ly-α). The energy dependence of the population of excited projectile levels was obtained for both targets. Furthermore, the experimental results were compared with theoretical calculations of nonradiative single- and double-electron capture based on the relativistic eikonal approximation and the independent-electron approximation. These findings provide valuable insights into the magnetic-sublevel population and *n*-resolved state-selective population of excited states produced in relativistic collisions of highly charged heavy ions with multi-electron atoms.

## 1 Introduction

Ion-atom collisions produce characteristic radiation through electron transfer, ionization, and excitation processes [1,2]. In highly charged heavy ions, radiative transitions provide critical insights into the effects of strong Coulomb fields on electronic structure [3]. For high-Z few-electron systems, transition energies and rates are particularly sensitive to relativistic and quantum electrodynamics (QED) corrections [4], making X-ray spectroscopy a powerful tool for testing strong-field QED predictions [5].

One of the most notable examples is the Ly-$\alpha_1$ transition ($2p_{3/2}$→$1s_{1/2}$) in hydrogen-like heavy ions, where the ground-state Lamb shift can be precisely deduced from the measured centroid energies of this transition [6]. The photon angular distribution of radiative transitions exhibits increased sensitivity to magnetic and retardation effects compared to total decay rates. A significant alteration in the photon angular distribution of the Ly-$\alpha_1$ transition arises from the interference between the dominant electric dipole $E1$ ($\Gamma_{E1} \propto Z^4$) and the much weaker magnetic quadrupole $M2$ ($\Gamma_{M2} \propto Z^8$) transition amplitudes during the decay of the $2p_{3/2}$ level in aligned hydrogen-like heavy ions. This provides observable evidence for multipole mixing [7,8]. Furthermore, the angular distribution and polarization of this Ly-$\alpha_1$ line reveal insights into the preferential population of magnetic-sublevels in high-Z ions. For the $2s_{1/2}$ decay in high-Z one-electron ions, relativistic effects become evident through the enhanced importance of $M1$ magnetic transitions, contrasting with the dominant $2E1$ decay observed at lower Z [9].

For He-like heavy ions, the K$\alpha_1$ and K$\alpha_2$ lines are possible separated through experimental procedures. Each transition line comprises two distinct components. Specifically, the K$\alpha_1$ line arises from the $^1P_1(E1)$ and $^3P_2(M2)$ states, while the K$\alpha_2$ line originates from the $^3S_1(M1)$ and $^3P_1(E1)$ states. Additionally, the continuous spectrum resulting from the $2E1$ decay of the $^1S_0$ level may be influenced by contributions from the $E1M1$ decay of the $^3P_0$ state [10].

In high-energy collisions of highly charged heavy ions with atoms, the precise spectroscopy of photons emitted by fast-moving projectiles is usually affected by the Doppler effect. This introduces a difference in the transition energies between the projectile frame and the laboratory frame due to energy Doppler shift at different observation angles and different collision energies. The energy of photons in the projectile frame $E_{\text{proj}}$ resulting from reaction processes (e.g., electron capture) corresponds to the measured energy in the laboratory frame $E_{\text{lab}}(\theta_{\text{lab}})$ according to the formula [11]

$$E_{\text{proj}} = E_{\text{lab}}(\theta_{\text{lab}}) \cdot \gamma \cdot (1 - \beta \cos \theta_{\text{lab}}), \qquad (1)$$

where $\theta_{\text{lab}}$ is the observation angle in the laboratory frame with respect to the beam axis, $\beta = \upsilon_{\text{proj}}/c$ denotes the projectile velocity $\upsilon_{\text{proj}}$ in units of the speed of light $c$, and $\gamma = 1/\sqrt{1-\beta^2}$ is the Lorentz factor. Concurrently, radiation emitted by the target atoms (resulting from the relaxation of excited states produced by charge transfer, excitation, or ionization) is unaffected by the Doppler shift.

Doppler-shifted X-ray measurements provide a powerful method for resolving closely spaced energy levels. For example, the Ly-$\alpha_1$ and Ly-$\alpha_2$ transitions in H-like xenon ions (separated by 427 eV) can be effectively resolved by selecting appropriate observation angles and collision energies to exploit the Doppler shift effect [12]. However, transitions characterized by extremely small energy separations, such as the 74 eV gap between the $^1P_1$ and $^3P_2$ states (components of the K$\alpha_1$ line in He-like uranium [13]) or the 76 eV difference between Ly-$\alpha_2$ and M1 transitions in H-like uranium [12], cannot be adequately distinguished using conventional germanium detectors due to their limited energy resolution. Recent advances in cryogenic calorimeter detectors have successfully distinguished the spectral lines of K$\alpha_1$, M2, K$\alpha_2$, and M1 transitions in $Xe^{52+*}$ ions produced in 50 MeV/u $Xe^{54+}$→Xe collisions [14] and $U^{90+*}$ generated in 10.225 MeV/u $U^{91+}$→$e^-$ collisions [15].

For symmetric heavy-ion collisions, the Doppler-shifted X-ray spectra of projectile ions may overlap with K X-rays from target ionization due to their similar energy ranges, particularly at intermediate observation angles. X-ray measurements at appropriately selected forward or backward angles can effectively avoid this spectral mixture (e.g., in $Xe^{54+}$→Xe collisions) [16].

The population of excited states via electron capture can be studied by observing the radiative decay of the excited levels to the ground state in high-Z few-electron systems. Particular attention should be given to the measurements of electron capture cross sections, which allow selective observation of final states. For heavy ions ($Z \geq 54$), their large subshell splitting allows such exclusive cross sections to be experimentally determined by measuring the X-ray emission associated with electron capture. This enables a more detailed testing of atomic reaction dynamics compared to measurement of total cross sections.

For electron capture from neutral target atoms into fast highly stripped projectiles, two mechanisms of radiative electron capture (REC) and non-radiative electron capture process (NRC) are often considered [17]. In NRC, a bound target electron is transferred to a bound state of the projectile, and the excess energy and momentum are shared between the target and the projectile [1]. NRC often competes with the REC process, where energy and momentum conservations are ensured through single photon emission [17]. As projectile energy decreases and target atomic number increases, NRC becomes increasingly significant compared to REC [1,17].

In addition to detailed studies of transition energies and probabilities in highly charged heavy ions, measurements of the angular distributions of characteristic X-ray photons provide important insights into the structure and dynamics of high-Z ions. One advantage of angle-resolved X-ray studies is their higher sensitivity to magnetic and retardation effects compared to the total cross section, which is obtained by integrating over all emission angles [7,17-20].

In recent years, new experiments have been conducted at the Heavy Ion Research Facility at Lanzhou-Cooling Storage Ring (HIRFL-CSR) [21,22] to measure transitions in few-electrons high-Z ions. These experiments [16,23-25] studied the population of the excited ionic states produced by nonradiative capture of target electron by bare heavy ions. Theoretically, the NRC process in fast collisions of highly charged ions with atoms has been explained using the relativistic eikonal approximation [26-28] and the two-center coupled-channel method [29,30], while the nonradiative double-electron capture (NRDC) process can be calculated using the many-body classical-trajectory Monte Carlo [31] and the independent-electron approximation [32-34].

Examples of x-ray spectra include the Ly-$\alpha_1$($2p_{3/2}\rightarrow 1s_{1/2}$) transition of $Xe^{53+}$ following NRC into the $2p_{3/2}$ state of initially bare $Xe^{54+}$ ions and K$\alpha_1$($^1P_1$, $^3P_2\rightarrow {}^1S_0$) radiative decays of $Xe^{52+}$ following NRDC into $^1P_1$, $^3P_2$ states. These experiments were conducted in collisions of $Xe^{54+}$ with Kr and Xe gas targets at projectile energies ranging from 95 to 197 MeV/u [16,23-25]. As the projectile energy increases, the angular distribution of the Ly-$\alpha_1$ transition changes from anisotropic to nearly isotropic emission [24,25]. Specifically, the magnetic-substate population ratio $\sigma\left(\frac{3}{2},\pm\frac{1}{2}\right)/\sigma\left(\frac{3}{2},\pm\frac{3}{2}\right)$ decreases rapidly from more than two to nearly one as the projectile energy increases. Additionally, a distinctly different energy dependence of the magnetic-substate population is observed compared to the results of the REC mechanism [24]. Furthermore, the measured total intensity ratio of the Ly-$\alpha_1$ and Ly-$\alpha_2$(+M1) radiations indicates that target electrons are preferentially captured to the $2p_{1/2}$ and $2s_{1/2}$ subshells rather than the $2p_{3/2}$ one [24,25]. When the projectile energy decreases to 50 MeV/u, the $2p_{3/2}$ population of $Xe^{53+}$ following NRC is significantly larger than the combined population of the $2p_{1/2}$ and $2s_{1/2}$ subshells for the Xe target [14]. At present, the alignment behaviors of the Ly-$\alpha_1$ transition from NRC and the K$\alpha_1$ radiation from NRDC in relativistic heavy ion collisions do not precisely describe yet due to the lack of an applicable theory [24,25,35].

Electron capture is one of the most important processes in the charge-changing mechanisms during collisions of few-electron heavy ions with atoms. This study allows for precise predictions of charge-state distributions and the lifetime of stored ion beams. Importantly, the measured results are importance for the operation of heavy-ion storage rings and accelerators, as well as for applications in nuclear physics experiments. Furthermore, electron capture cross sections serve as fundamental input parameters in various models for plasma and astrophysics [36,37]. Additionally, experimental results on electron capture provide valuable tests for the theory of atomic collision dynamics [38].

In this paper, we report Doppler-shifted X-rays produced in electron-capture experiments at the HIRFL-CSR storage ring during collisions of 95–197 MeV/u $Xe^{54+}$ ions with Kr and Xe gas atoms. The measured angular distributions of the K$\alpha_1$(+M2) transition of $Xe^{52+}$ ions produced via the NRDC process are used to determine the anisotropy parameters. These parameters depend on the alignment parameters of the $\left(1s_{1/2}2p_{3/2}\right)_1$ and $\left(1s_{1/2}2p_{3/2}\right)_2$ states as well as on their relative populations. Additionally, the relative populations of the excited L, M, and N-shell levels of $Xe^{52+}$ and $Xe^{53+}$ (resulting from the NRC process) were deduced from the intensity ratios of I[Kα$_1$(+M2)]/I[Ly-α$_2$(+M1)], I(Ly-β)/I(Ly-α), I(Ly-γ)/I(Ly-α), and I(Kα)/I(Ly-α). These experimental findings may provide insights into the selective population of lowly-excited states in collisions of highly charged high-Z ions with atoms, where projectile velocities are comparable to the orbital velocities of inner-shell electrons of both the projectile ions and target atoms. Furthermore, these results can help advance the development of applicable relativistic NRC and NRDC methods.

## 2 Experiment

The experiment was conducted at the experimental ring (CSRe), which is a component of the HIRFL-CSR complex at the Institute of Modern Physics, Chinese Academy of Sciences in Lanzhou, China. Xenon ions with low charge states were generated by a superconducting electron cyclo ato tron resonance ion source, preaccelerated in a sector-focusing cyclotron, and subsequently injected into the main ring of the HIRFL-CSR. These xenon ions were accumulated, cooled, and further accelerated to the desired beam energies. Prior to injection into the CSRe, the ion beam passed through a carbon stripper foil of 45 mg/cm$^2$ thickness to produce the bare $Xe^{54+}$ ions. The final energies of 95, 146, and 197 MeV/u within the CSRe were measured and monitored by revolution frequencies and Schottky noise analysis. The energy range corresponds to the relativistic $\beta$ values from 0.42 to 0.56. The precise measurement of ion velocities is well suited for atomic-structure investigations of very heavy few-electron systems. Typical beam intensities of several $10^7$ bare xenon ions were stored. In the CSRe, electron cooling was employed to achieve a relative momentum spread in the order of $\Delta p/p \sim 10^{-5}$ for the $Xe^{54+}$ ions.

The internal gas target [39] of the CSRe was activated following the completion of the cooling cycle. Xenon and krypton gaseous targets with densities ranging from $10^{12}$ and $10^{13}$ atoms/cm$^2$ were employed. The overlap between the ion beam and the target was optimized by scanning the ion beam position across the perpendicular gas jet, while continuously monitoring the counting rate of two photomultipliers. Within the energy range of interest, the most relevant charge-exchange processes occurring in ion-atom collisions at the gas target are REC and NRC. Consequently, the single-collision condition was well satisfied by the total capture cross section of approximately $10^{-20}$ cm$^2$ in the present experiments [1,17,26-28].

X-rays emitted from the interaction zone between the ion beam and the gas target were measured using an array of six X-ray detectors. Four of these detectors were high-purity germanium (HPGe) detectors, positioned at observation angles of 35°, 60°, 90°, and 120° with respect to the ion beam axis. These four detectors were located at respective distances of 500, 300, 270, and 360 mm from the interaction point. Additionally, two lithium-drifted silicon (Si(Li)) detectors were placed at angles of 90° and 145°, with respective distances of 270 and 500 mm. All x-ray detectors were mounted behind 100-μm beryllium windows, which separated them from the ultrahigh-vacuum system of the interaction chamber. The detectors were shielded by lead and brass assemblies and collimated by 8-mm diameter holes. The solid angles ($\Delta\Omega/4\pi$) covered by the individual detectors ranged from approximately $10^{-6}$ to $10^{-4}$. The detectors were calibrated using radioactive sources of $^{55}$Fe, $^{133}$Ba, $^{152}$Eu, and $^{241}$Am before and after the experiment. The typical energy resolution (i.e., full width at half maximum) of the used detectors was approximately 300 eV at around 30 keV [16,25]. Further details regarding the experimental setup have been previously described in Refs. [16,23-25].

## 3 Data analysis and results

Figure 1 shows examples of X-ray spectra recorded by the detector positioned at 145° in collisions of 95 MeV/u $Xe^{54+}$ ions with Kr and Xe targets. The spectra reveal negligible X-ray radiation corresponding to REC into the empty *K*- and *L*-shells of the projectile. Instead, the observed spectra are dominated by the Lyman series of down-charged projectile ions and decay lines arising from target ionization. This observation indicates that the projectile Lyman series are produced primarily via the NRC mechanism. As supported by theoretical calculations, the shell-resolved NRC cross sections are significantly larger than the REC cross sections, as described in Refs. [16,25]. It should be noted that NRC directly into the projectile ground state does not result in subsequent X-ray emission.

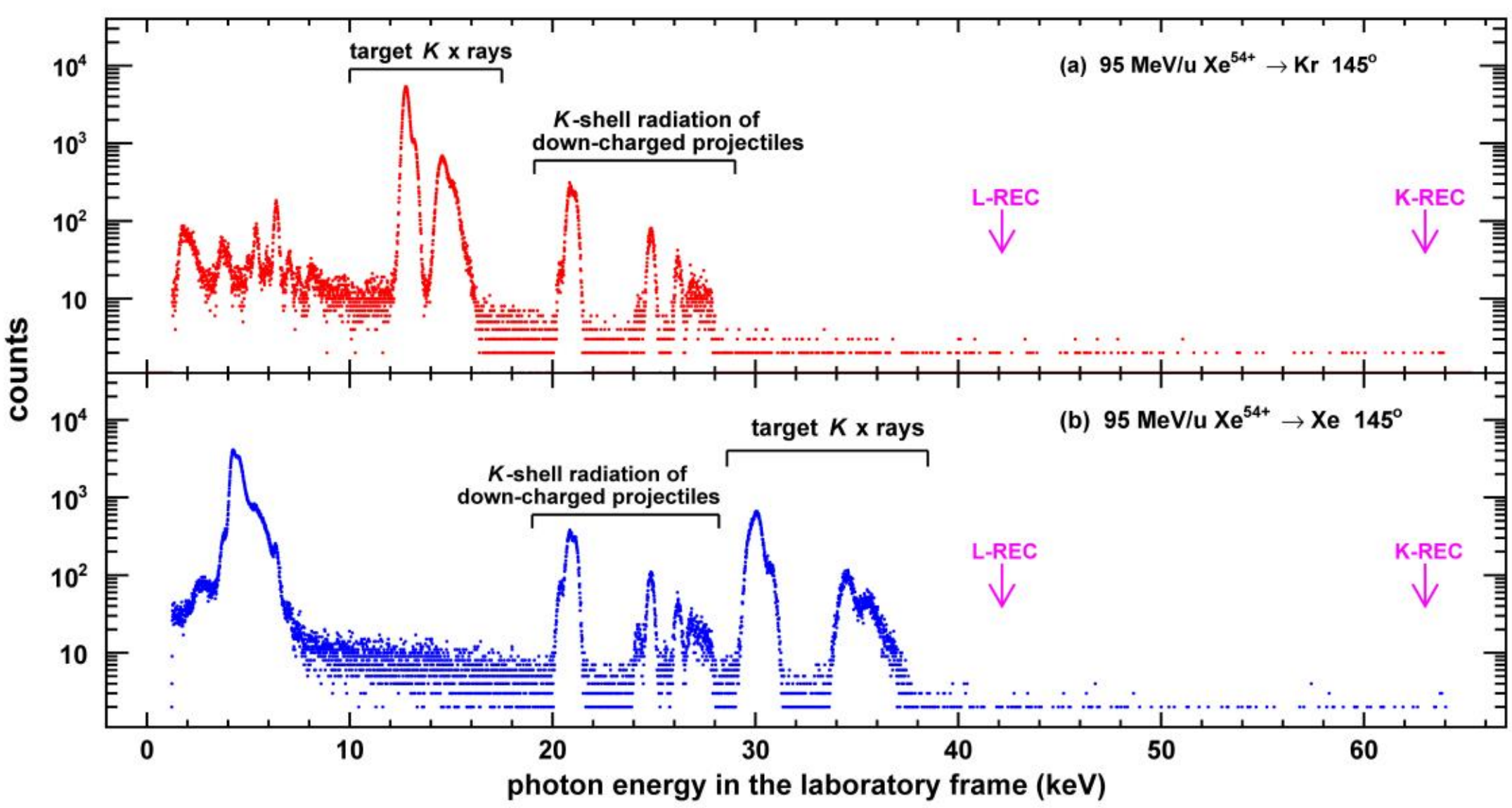

Figure 1. X-ray spectra resulting from collisions of 95 MeV/u $Xe^{54+}$ ions with (a) Kr and (b) Xe targets were recorded at 145° by a Si(Li) detector and not corrected for detection efficiency. The *K*- and *L*-REC transitions are negligible compared to the Lyman series of the down-charged projectiles.

Table 1. Characteristic transitions of $Xe^{53+}$ [12] and $Xe^{52+}$ [13] ions corresponding to the spectra presented in Figures 3-6.

| Ion | Transition | Type | Energy (keV) | Transition Probability ($s^{-1}$) | Group | Note |
|---|---|---|---|---|---|---|
| $Xe^{53+}$ | $2p_{3/2} \rightarrow 1s_{1/2}$ | Ly-α1 | 31.284 | $5.1 \times 10^{15}$ | I | Electric dipole |
| | $2p_{1/2} \rightarrow 1s_{1/2}$ | Ly-α2 | 30.856 | $5.4 \times 10^{15}$ | II | Electric dipole |
| | $2s_{1/2} \rightarrow 1s_{1/2}$ | $M1$ | 30.863 | $6.3 \times 10^{11}$ | | Magnetic dipole |
| $Xe^{52+}$ | $(1s_{1/2}2p_{3/2})_1 \rightarrow 1s^2$ | $K\alpha_1$ | 30.630 | $6.8 \times 10^{15}$ | III | Electric dipole<br>$^1P_1 \rightarrow {}^1S_0$ |
| | $(1s_{1/2}2p_{3/2})_2 \rightarrow 1s^2$ | $M2$ | 30.594 | $2.6 \times 10^{12}$ | | Magnetic quadrupole<br>$^3P_2 \rightarrow {}^1S_0$ |
| | $(1s_{1/2}2p_{1/2})_1 \rightarrow 1s^2$ | $K\alpha_2$ | 30.206 | $3.0 \times 10^{15}$ | IV | Electric dipole<br>$^3P_1 \rightarrow {}^1S_0$ |
| | $(1s_{1/2}2s_{1/2})_1 \rightarrow 1s^2$ | $M1$ | 30.129 | $3.8 \times 10^{11}$ | | Magnetic dipole<br>$^3S_1 \rightarrow {}^1S_0$ |

Table 2. Doppler-shifted energies from projectile frame to laboratory frame for selected transitions.

| Projectile energy (MeV/u) | Transition | $E_{proj}$ (keV) | $E_{lab}$ (keV) | | | | |
|---|---|---|---|---|---|---|---|
| | | | 35° | 60° | 90° | 120° | 145° |
| 197 | Ly-α1 | 31.284 | 48.035 | 35.978 | 25.823 | 20.139 | 17.658 |
| | Ly-α2 | 30.856 | 47.378 | 35.485 | 25.470 | 19.863 | 17.416 |
| | K-REC | 149.367 | 229.344 | 171.776 | 123.292 | 96.153 | 84.307 |
| 146 | Ly-α1 | 31.284 | 45.974 | 36.123 | 27.045 | 21.613 | 19.157 |
| | Ly-α2 | 30.856 | 45.345 | 35.629 | 26.675 | 21.318 | 18.895 |
| | K-REC | 121.392 | 178.394 | 140.171 | 104.943 | 83.866 | 74.337 |
| 95 | Ly-α1 | 31.284 | 43.287 | 35.939 | 28.389 | 23.460 | 21.120 |
| | Ly-α2 | 30.856 | 42.695 | 35.447 | 28.000 | 23.139 | 20.831 |
| | K-REC | 93.415 | 129.256 | 107.314 | 84.770 | 70.053 | 63.065 |

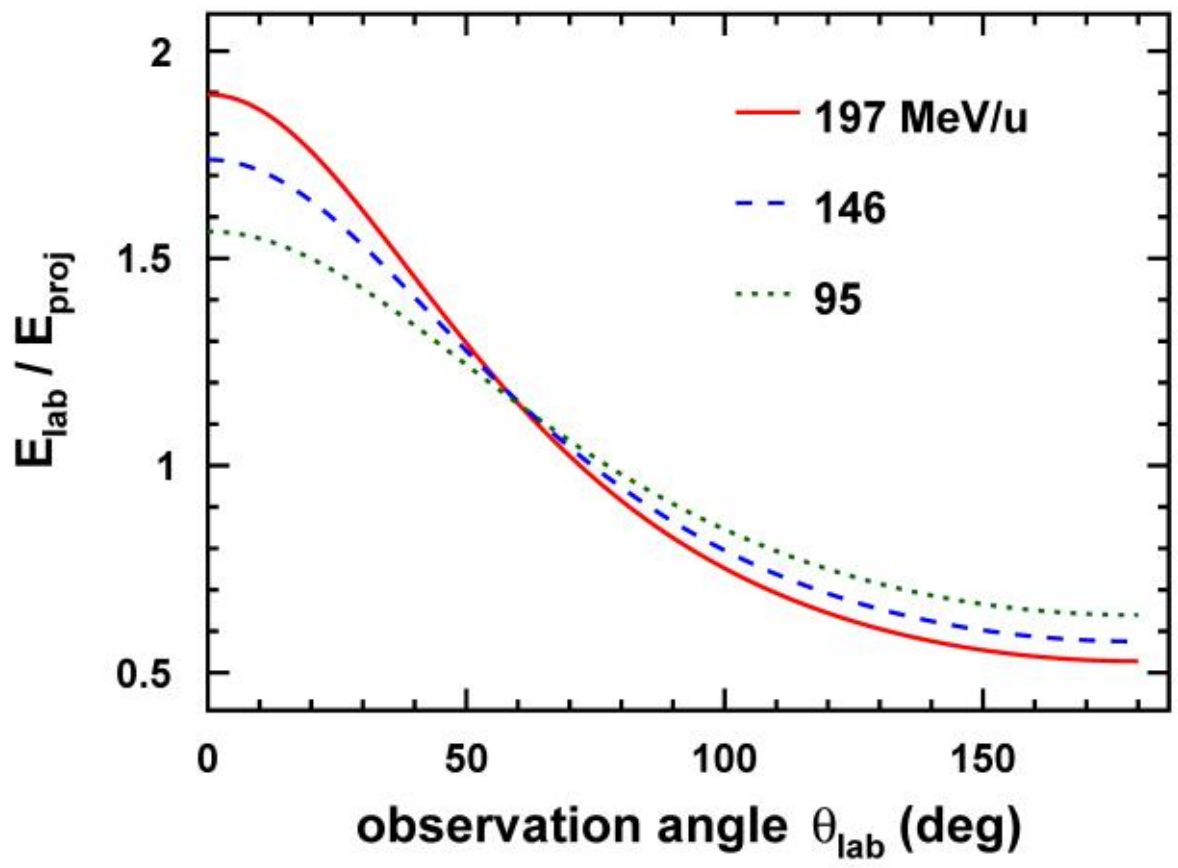

Figure 2. Doppler shift of transition energy from the projectile frame to the laboratory frame for projectile ions moving with reduced velocities of $\beta \approx 0.56$ (197 MeV/u), 0.50 (146 MeV/u), and 0.42 (95 MeV/u), as a function of observation angle.

In the following analysis, we focus on the X-ray spectra radiated by the down-charged projectile ions. To identify and distinguish the contributions of different radiation within these spectra, the primary transition energies for H- and He-like xenon ions are essential. The energies and transition probabilities of the Ly-α and Kα transitions, calculated in Refs. [12,13], are listed in Table 1. For the present experimental study, the laboratory-frame energies for Ly-α and K-REC transitions, corrected for the Doppler shift using Equation (1), are calculated and listed in Table 2. The ratio $E_{lab}/E_{proj}$ is plotted as function of observation angles for projectile energies of 95, 146, and 197 MeV/u in Figure 2.

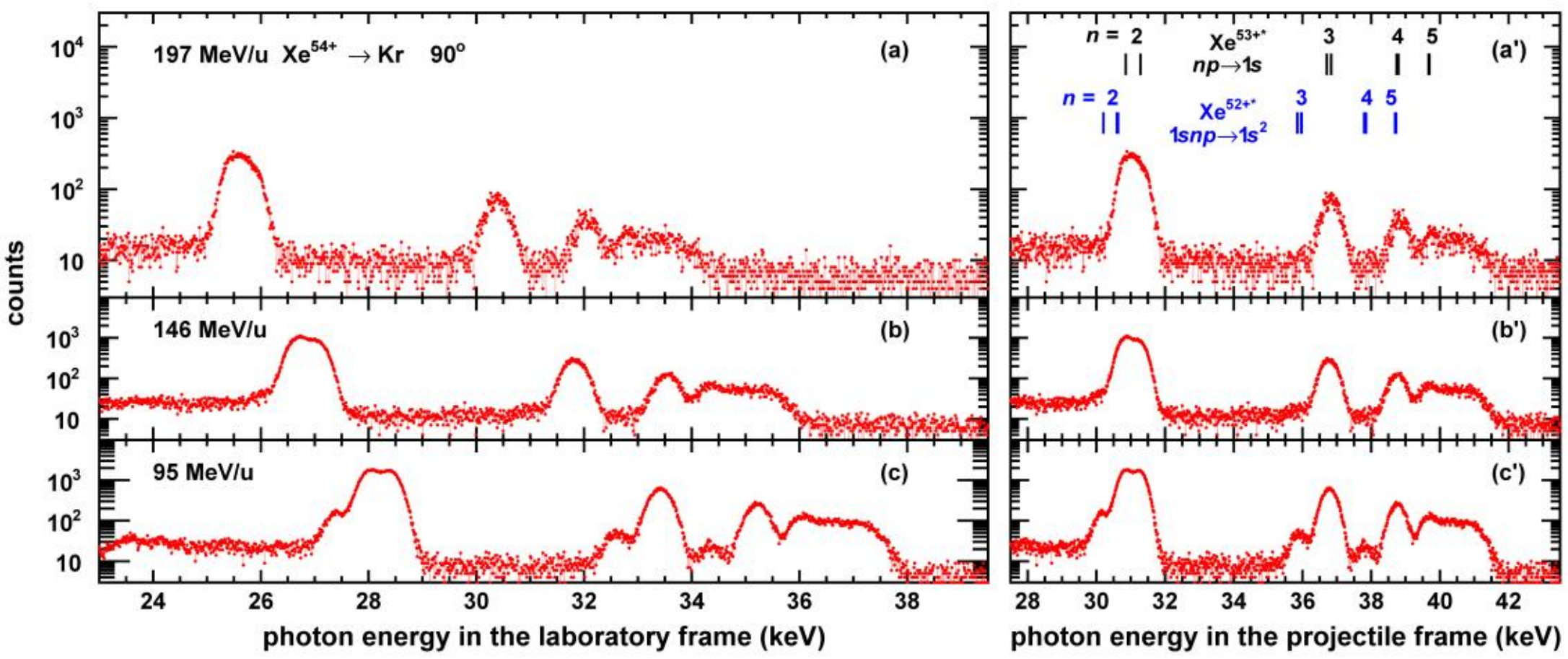

Figure 3. X-ray spectra resulting from $Xe^{54+} \rightarrow Kr$ collisions measured at 90° by an HPGe detector. The X-ray photons were emitted in the subsequent radiative decay of down-charged projectile ions following single- and double-electron capture into excited projectile states. The spectra are presented respectively in the laboratory (left panel) and projectile (right) frames for three projectile energies of (panels a and a') 197 MeV/u, (b and b') 146 MeV/u, and (c and c') 95 MeV/u. The transitions $1snp \rightarrow 1s^2$ ($n$=2, 3, 4, and 5) of $Xe^{52+*}$ and $np \rightarrow 1s$ of $Xe^{53+*}$ are indicated by vertical lines in the right panel. The spectra are not corrected for detection efficiency.

As an example, Figure 3 displays the X-rays emitted from the decay of $Xe^{53+}$ and $Xe^{52+}$ ions following single and double electron capture in collisions of 95, 146, and 197 MeV/u $Xe^{54+}$ projectiles with Kr targets, measured at the observation angle of 90° in the laboratory frame. The Doppler-corrected $np \rightarrow 1s$ and $1snp \rightarrow 1s^2$ transitions are also shown. The Ly-α, Ly-β, and Ly-γ transitions of $Xe^{53+}$ produced by the NRC process [16], are clearly resolved, while transitions from higher levels overlap. The $K$α, $K$β, and $K$γ transitions of $Xe^{52+}$ resulting from the NRDC process as detailed in Ref. [23] are also present in the spectra at the projectile energy of 95 MeV/u. These spectral lines are less distinct at 146 MeV/u and absent at the higher energy of 197 MeV/u.

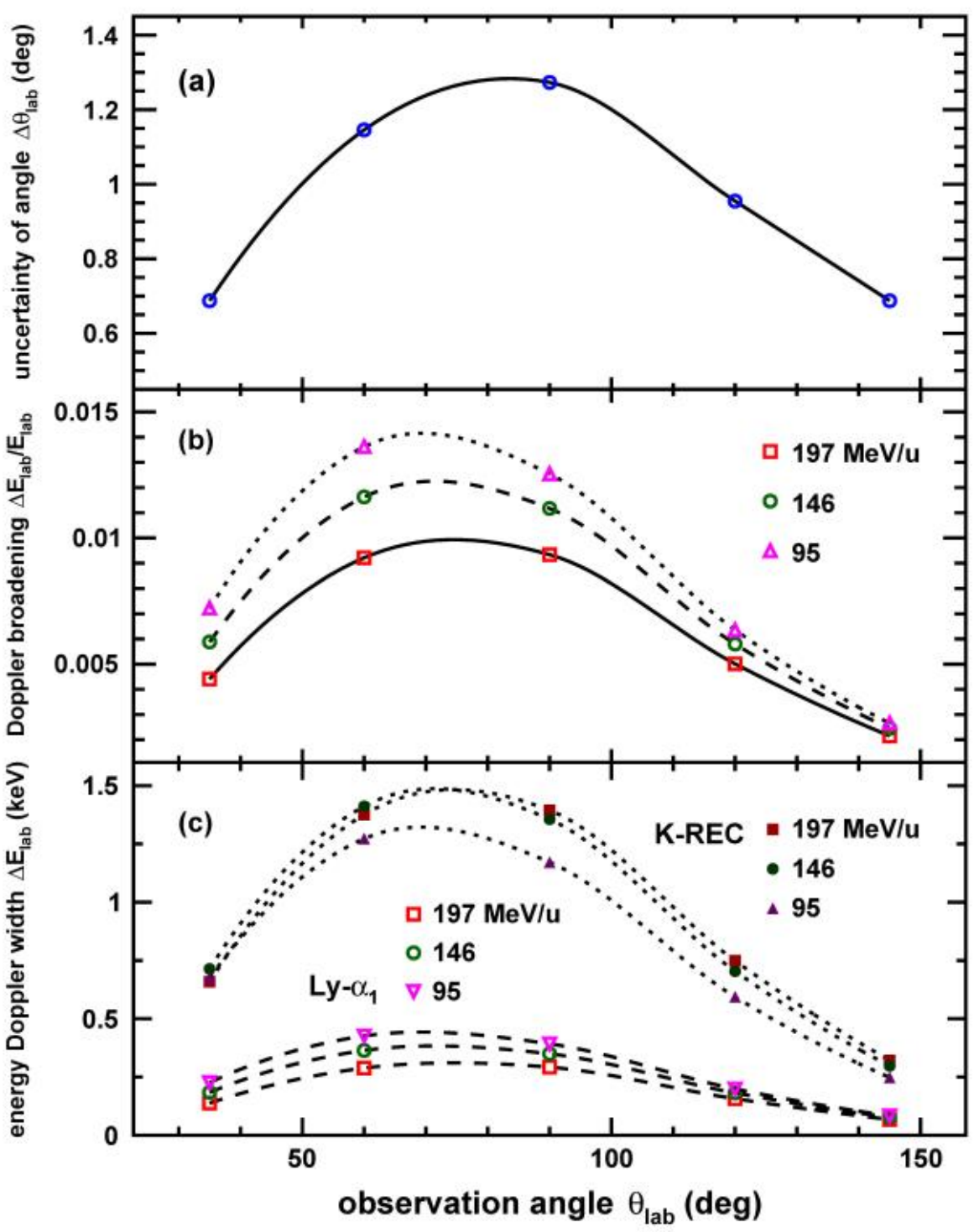

Figure 4. (a) Uncertainties in the observation angles $\Delta\theta_{\mathrm{lab}}$, (b) Doppler broadening $\Delta E_{\mathrm{lab}}/E_{\mathrm{lab}}$ at projectile energies of 197, 146, and 95 MeV/u for transitions in H-like and He-like xenon ions, and (c) energy Doppler width $\Delta E_{\mathrm{lab}}$ for the Ly-$\alpha_1$ transition and K-REC photons at observation angles of 35°, 60°, 90°, 120°, and 145°. These results were calculated by accounting for the diameter of the gas-jet target (about 4 mm) and the collimator aperture diameter (8 mm) of the detectors.

Another Doppler effect influencing the measured X-ray energy spectra is the energy broadening $\Delta E_{\mathrm{lab}}$ arising from the uncertainty in the observation angle $\Delta\theta_{\mathrm{lab}}$. This Doppler broadening can be calculated using

$$\frac{\Delta E_{\mathrm{lab}}}{E_{\mathrm{lab}}} = \frac{\beta \sin\theta_{\mathrm{lab}}}{1-\beta\cos\theta_{\mathrm{lab}}} \Delta\theta_{\mathrm{lab}}. \qquad (2)$$

Considering the diameter of the gas-jet target (about 4 mm) and the collimator aperture diameter (8 mm) of the detectors, uncertainties in observation angles for the present experiment are presented in Figure 4(a). The corresponding Doppler broadening $\frac{\Delta E_{\mathrm{lab}}}{E_{\mathrm{lab}}}$ and $\Delta E_{\mathrm{lab}}$ for the Ly-$\alpha_1$ transition of $Xe^{53+}$ and K-REC photons at projectile energies of 197, 146, and 95 MeV/u are also shown in Figure 4. It is noted that reduction of the angular width significantly decreases the broadening of the measured transition energies, particularly near observation angles 0° and 180°. The use of different collimators in front of the detectors can improve the separation of the different energies corresponding to the various transitions in the H- and He-like high-Z ions. However, apart from this improvement, collimating the detectors implies reduction of solid angle by reducing the detector active area and therefore of the reduction in the detection efficiency.

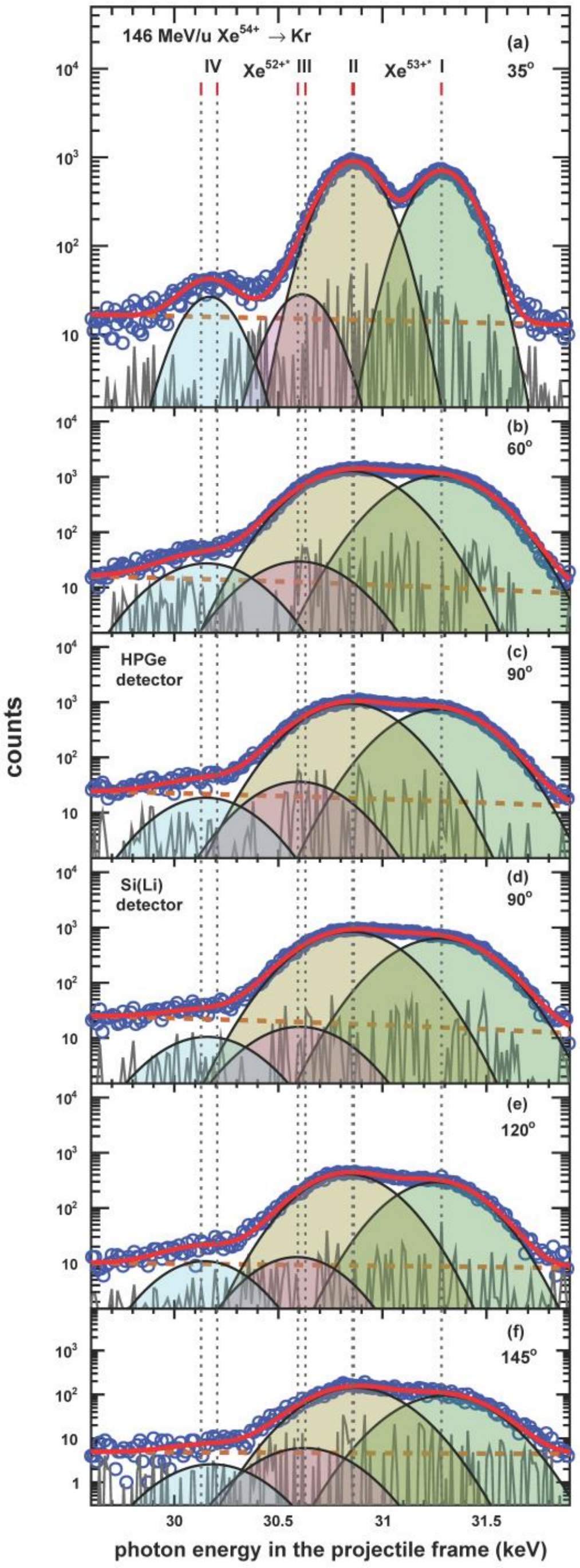

Figure 5. X-ray spectra from the ground-state transitions $1s2p \rightarrow 1s^2$ of $Xe^{52+}$ and $2p \rightarrow 1s$ of $Xe^{53+}$ produced in 146 MeV/u $Xe^{54+}$→Kr collisions. These spectra were recorded at (a) 35°, (b) 60°, (c) 90°, and (e) 120° by four HPGe detectors as well as at (d) 90° and (f) 145° using two Si(Li) detectors. The transitions resulting from the NRDC and NRC processes are labeled the same as in Figure 3. The corresponding transition energies for $Xe^{52+}$

[13] and $Xe^{53+}$ [12] listed in Table 1, are indicated by dashed vertical lines. Peak intensities are determined by a fitting procedure.

In the present experiment, the energy Doppler width is comparable to the intrinsic energy resolution of the detectors. However, the observed X-ray line width is determined by the Doppler broadening and the detector energy resolution. To simplify data analysis procedures, intensities of spectral transition lines (e.g., Kα and Ly-α) were determined by fitting Gaussian line profiles, incorporating detection efficiency correction.

The Ly-$\alpha_{1,2}$ and $M1$ transitions of $Xe^{53+}$ as well as the $K\alpha_{1,2}$, $M1$, and $M2$ transitions of $Xe^{52+}$ give rise to partially overlapping X-ray spectra in 146 MeV/u $Xe^{54+}$→Kr collisions at observation angles of 35°, 60°, 90°, 120°, and 145°, as shown in Figure 5. Intensities of these seven transitions are determined by four Gaussian peaks (i.e., $2p_{3/2} \to 1s_{1/2}$, $2p_{1/2} \to 1s_{1/2}$ mixed with $2s_{1/2} \to 1s_{1/2}$, $\left(1s_{1/2}2p_{3/2}\right)_1 \to 1s^2$ mixed with $\left(1s_{1/2}2p_{3/2}\right)_2 \to 1s^2$, and $\left(1s_{1/2}2p_{1/2}\right)_1 \to 1s^2$ mixed with $\left(1s_{1/2}2s_{1/2}\right)_1 \to 1s^2$) in the fitting procedure according to the transition energies and detector energy resolution [16]. The fitting peaks marked as I, II, III, and IV are listed in Table 1 and Figure 5. The intensities of these Gaussian peaks are adjustable fitting parameters. The width of these peaks is set to be a common free parameter because it is defined by detector resolution within this energy region. Simultaneously, spacing ratios among these peaks are fixed according to the corresponding transition energies for minimizing fitting uncertainty. A linear background is also incorporated. The fitting uncertainties are estimated from the fitting residuals. The obtained intensities of these peaks are then corrected for the energy-dependent detection efficiency (with an error of 3% for this correction).

Additionally, the peak intensities of the Ly-β and Ly-γ transitions were determined by similar fitting procedures. Consequently, the intensity ratios of $I(\text{Ly-}\alpha_1)/I(\text{Ly-}\alpha_2 + M1)$, $I(\text{Ly-}\beta)/I(\text{Ly-}\alpha)$, $I(\text{Ly-}\gamma)/I(\text{Ly-}\alpha)$, $I(\text{K}\alpha_1 + M2)/I(\text{Ly-}\alpha_2 + M1)$, and $I(\text{K}\alpha)/I(\text{Ly-}\alpha)$ were derived. It is important to note that other higher-multipole radiative transitions from the $1sns$ ($n$≥2) and $1snd$ ($n$≥3) states of $Xe^{52+}$ as well as the $ns$ ($n$≥2) and $nd$ ($n$≥3) states of $Xe^{53+}$ may overlap with the observed spectral lines of the down-charged projectile ions as shown in Figures 3 and 5. In the present experiment, the mixture of these transitions cannot be resolved due to limited energy resolution of the used photon detectors and additional Doppler broadening induced by the fast-moving projectiles. Therefore, the measured transitions $1snp \to 1s^2$ ($n$=2, 3, and 4) correspond to ground-state decays originating from the *KL*, *KM*, and *KN* shells of $Xe^{52+}$. Similarly, the transitions $np \to 1\text{s}$ ($n$=2, 3, and 4) represent decays from the *L*, *M*, and *N* shells of $Xe^{53+}$.

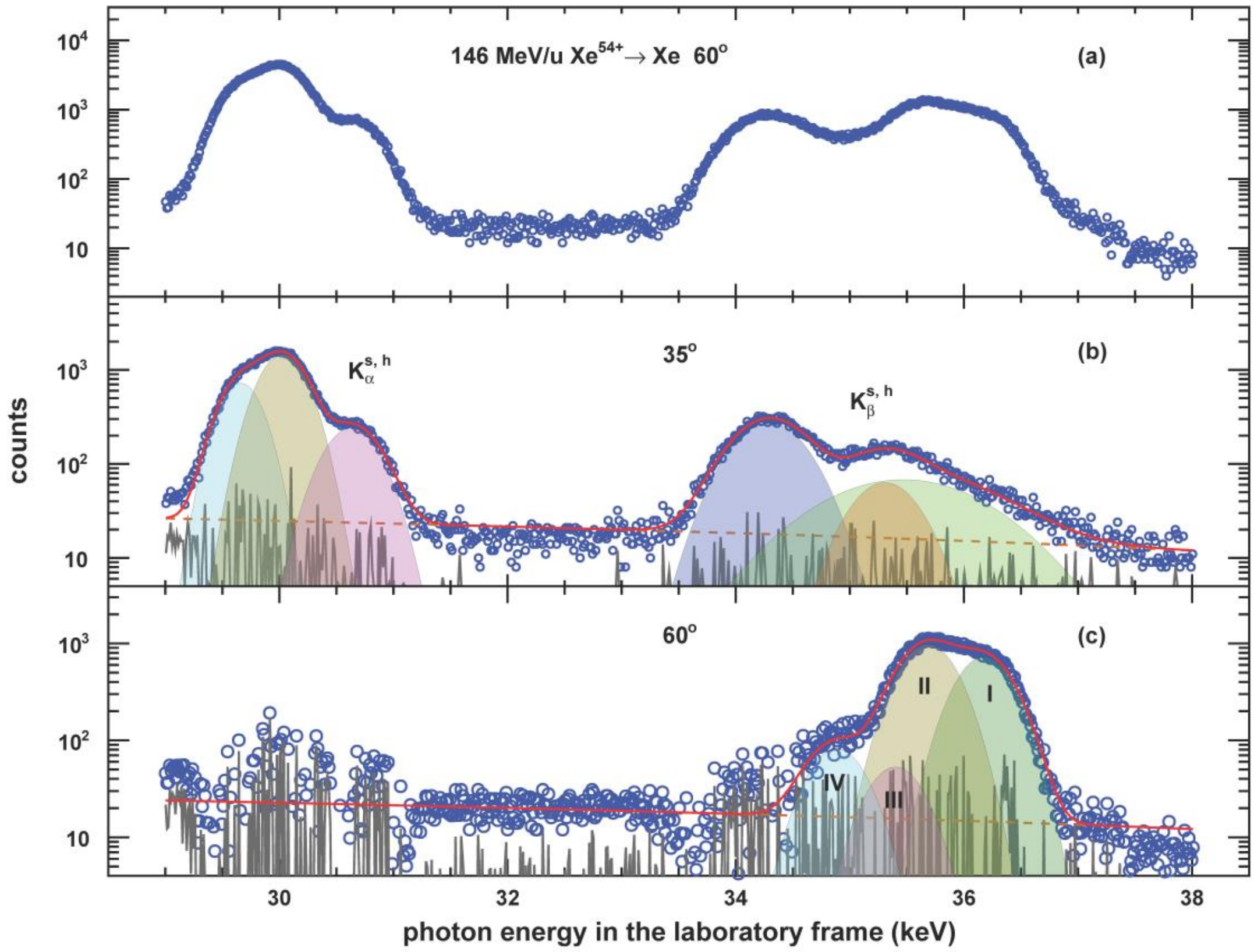

Figure. 6. Determination of transition intensities for down-charged projectiles when X-rays emitted from projectiles overlap with the *K* X-rays resulting from target ionization in 146 MeV/u $Xe^{54+}$→Xe collisions. (a) Spectrum measured by an HPGe detector at 60°. Decay lines from the down-charged projectiles are superimposed on deexcitation lines following the *K*-shell ionization of target Xe. (b) Spectrum of Xe *K*-shell ionization obtained by the same type HPGe detector at 35°. Doppler-shifted transition energies for projectile X-rays exceed 43 keV at this angle. (c) X-ray spectrum of the down-charged projectiles at 60° after subtraction of the Xe *K*-shell ionization. The corresponding peak fitting procedure by using the above-mentioned method is also given. The peaks marked as I, II, III, and IV are listed in Table 1.

For symmetric $Xe^{54+}$→Xe collisions at observation angles of 60°, 90°, and 120° in the present work, the Doppler-shifted X-ray spectra of down-charged projectile ions may overlap with the *K* X-rays arising from Xe ionization. We use the spectrum obtained for 146 MeV/u $Xe^{54+}$→Xe collisions at 60° as an illustrative example. As shown in Figure 6(a), the transitions of $Xe^{53+}$ and $Xe^{52+}$ ions overlap with the $K\beta$ satellite and hyper-satellite lines of target ions. Target *K* X-rays observed at 35° by another detector with the same type are displayed in Figure 6(b). At the angle of 35°, the projectile X-rays were Doppler-shifted to energies above 43 keV. Since the target *K* X-rays were observed to be isotropic in the present experiment [1,16], the shape of the target spectrum was fitted by a group of six Gaussian peaks and a linear background as shown in Figure 6(b). The shape was then employed as a single fitting function to subtract the contribution of the target X-rays from the spectrum measured at 60°. The subtracted spectrum was subsequently fitted to determine the intensities of the projectile transitions, as shown in Figure 6(c). The uncertainty associated with the subtraction procedure was estimated based on the fitting residuals. Similar subtraction procedures were applied to resolve overlapping spectra at other observation angles.

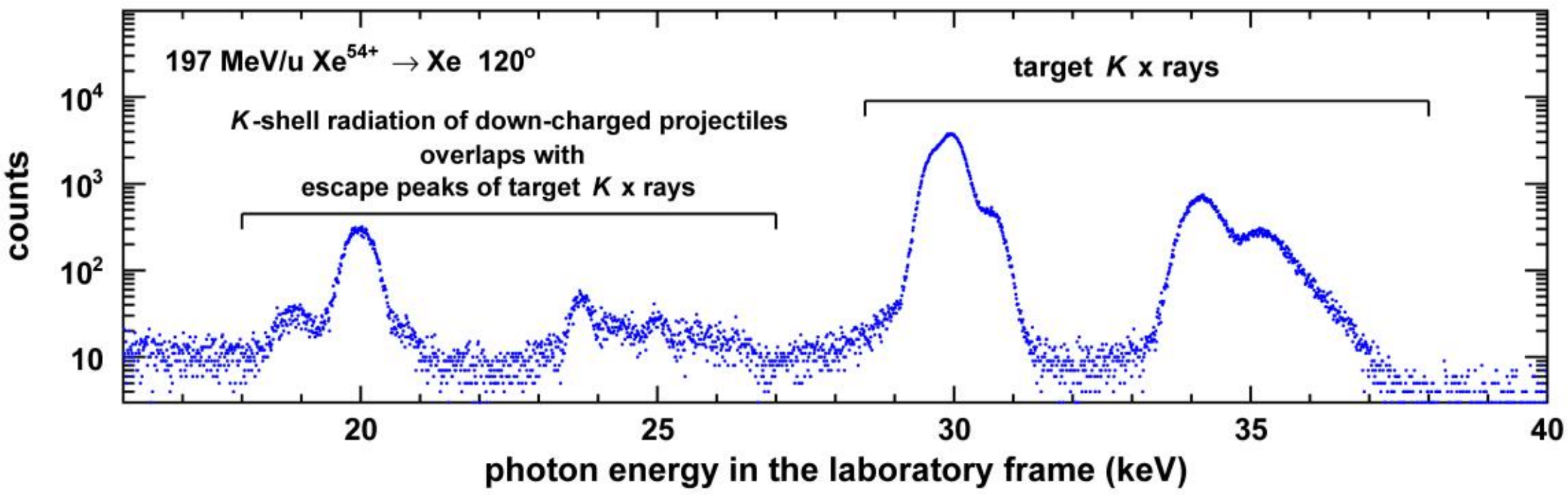

Figure 7. X-ray spectrum from 197 MeV/u $Xe^{54+}$→Xe collisions recorded by an HPGe detector at 120° and are uncorrected for detection efficiency. *K*-shell transition lines of down-charged projectiles were contaminated by escape peaks of the intense target *K* X-rays.

It is important to note that escape peaks resulting from the germanium *K* radiation leaving the detector crystal may overlap with Doppler-shifted X-rays of down-charged projectiles produced by the NRC process in the present work. For instance, Figure 7 displays the X-ray spectrum measured at 120° in 197 MeV/u $Xe^{54+}$→Xe collisions, where the target radiation intensity of is significantly exceeds that of the projectile radiation. Consequently, escape peaks originating from intense target *K* X-rays contaminate the projectile transition lines (as seen in Figure 7), which makes spectral separation difficult.

Moreover, the X-ray detectors were shielded with lead to reduce background radiation. Ionization of *K*-shell electrons in the lead atoms by ambient radiation may produce fluorescence radiation near 74 keV for Pb-Kα and 85 keV for Pb-Kβ in the X-ray spectra. Meanwhile, Doppler-shifted transitions from down-charged xenon ions measured across three energies and five angles remain below 66 keV. The energy separation prevents overlap between projectile X-rays and Pb Kα and Kβ lines in the present experiment.

The population distribution of excited states in H-like xenon ions resulting from NRC and He-like xenon ions produced by NRDC can be deduced from precision measurement of their radiative decay to the ground state.

As seen in Figure 5, the emission lines of Ly-$\alpha_1$, Ly-$\alpha_2$(+M1), K$\alpha_1$(+M2), and K$\alpha_2$(+M1) are observed. The single-electron capture process is found to dominate over the double-electron capture. Changes in the relative intensities between the Ly-$\alpha_1$ and Ly-$\alpha_2$(+M1) components, as well as between the K$\alpha_1$(+M2) and K$\alpha_2$(+M1) transitions, can be noticed. Furthermore, the energy dependence and target dependence of the population of excited projectile levels can be revealed by combining with the results at other energies (95 and 146 MeV/u) and with the Xe target. These results serve as a highly sensitive probe for theoretical methods describing collision dynamics and the structure of high-Z ions [7,17].

The Ly-$\alpha_2$(+M1) transition exhibits isotropic emission as discussed in Refs. [17,40]. Consequently, it serves as an ideal reference for measuring potential anisotropy in the neighboring Ly-$\alpha_1$ and K$\alpha$ transitions. By using the Ly-$\alpha_2$(+M1) transition for normalization, systematic uncertainties associated with corrections of solid angle and detection efficiency are substantially reduced or canceled out. The technique has been applied in previous studies to analyze the angular distributions of the intensity ratio $I(\text{Ly-}\alpha_1)/I(\text{Ly-}\alpha_2 + M1)$ of H-like xenon ions produced by the NRC process in relativistic ion-atom collisions [24,25]. The corresponding anisotropy parameters of the Ly-$\alpha_1$ transition, as well as the alignment parameters and the magnetic sublevel population of the excited $Xe^{53+}(2p_{3/2})$ state were directly deduced. The procedure from the emission pattern (dependence on the transition) to the population of corresponding magnetic sublevels is described in detail in Refs. [17,24,25]. Moreover, the experimental relative cross sections for NRC populating different total angular momentum states within the *L*-shell of H-like xenon ions were also determined from the observed intensities of the projectile Ly-$\alpha_1$ and Ly-$\alpha_2$(+M1) radiation.

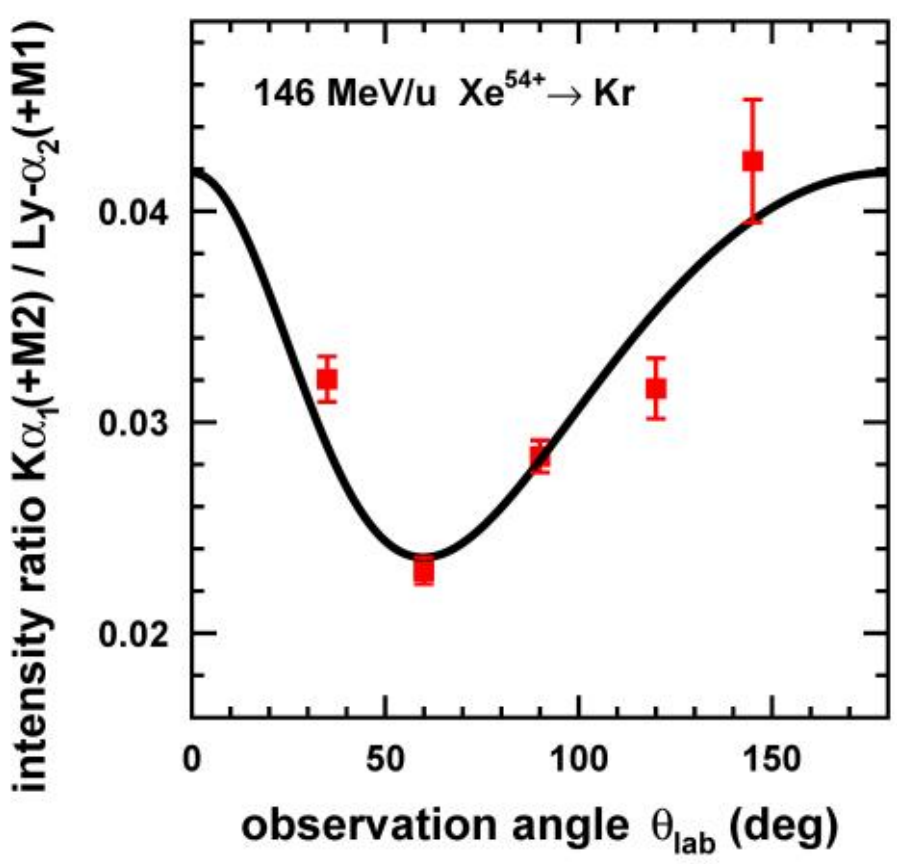

Figure 8. Angular distribution of the K$\alpha_1$(+*M*2) radiation in $Xe^{52+}$ ions produced from 146 MeV/u $Xe^{54+}$→Kr collisions. The intensity of K$\alpha_1$(+*M*2) is normalized by the isotropic Ly-$\alpha_2$(+*M*1) transition. The solid line represents the fitting of the experimental data according to the formula of the angular distribution of the intensity ratio between K$\alpha_1$(+M2) and Ly-$\alpha_2$(+M1) transitions as described in Ref. [17,35]. The error bars are estimated by the statistics of data.

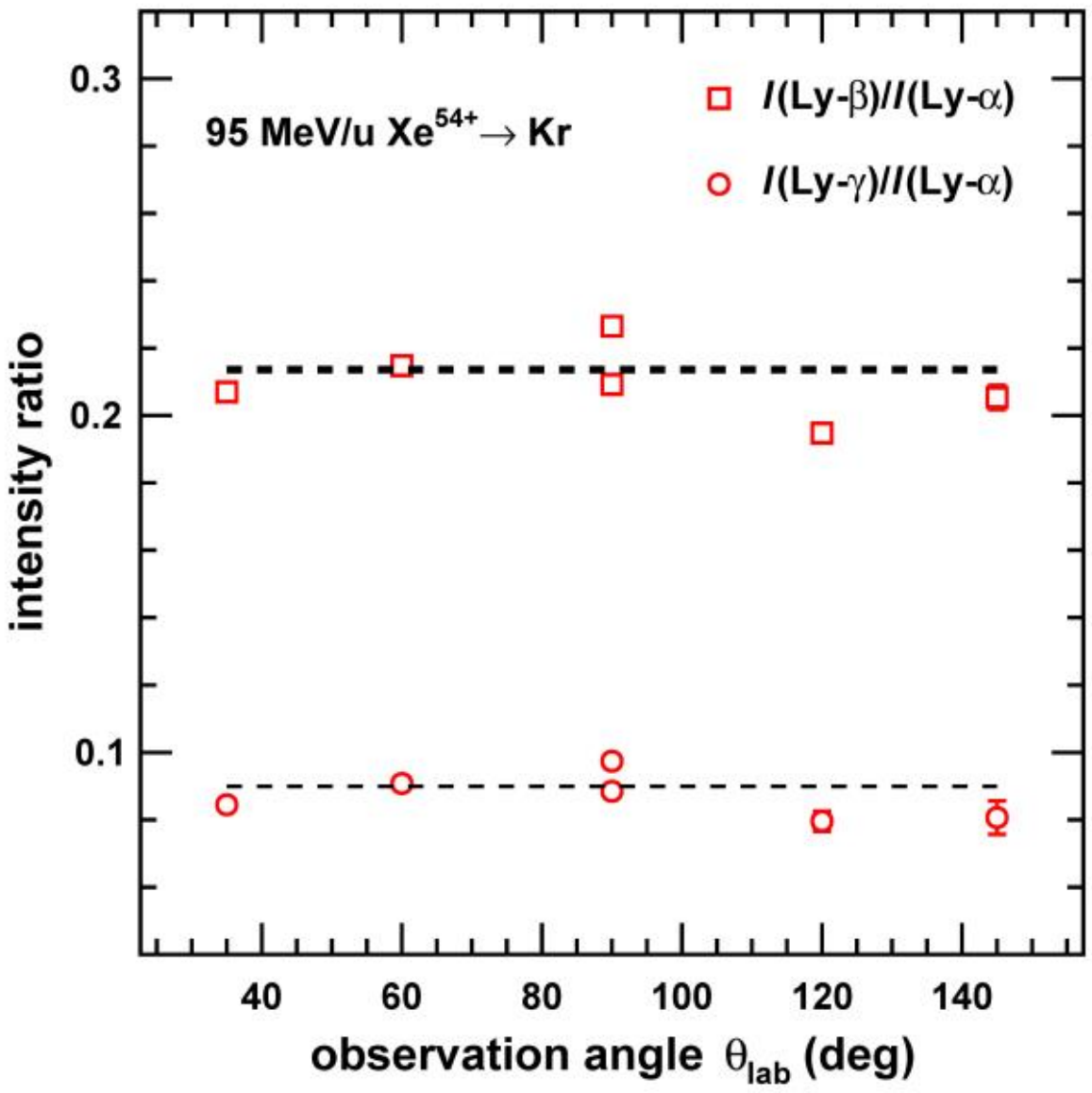

FIG. 9. Angular dependences of the intensity ratios $I(\text{Ly-}\beta)/I(\text{Ly-}\alpha)$ and $I(\text{Ly-}\gamma)/I(\text{Ly-}\alpha)$ for the $Xe^{53+*}$ ions in 95 MeV/u $Xe^{54+}$→Kr collisions. The error bars are estimated by the statistics of data. The weighted average values of the intensity ratios (i.e. the overall intensity ratios) are illustrated by dash lines.

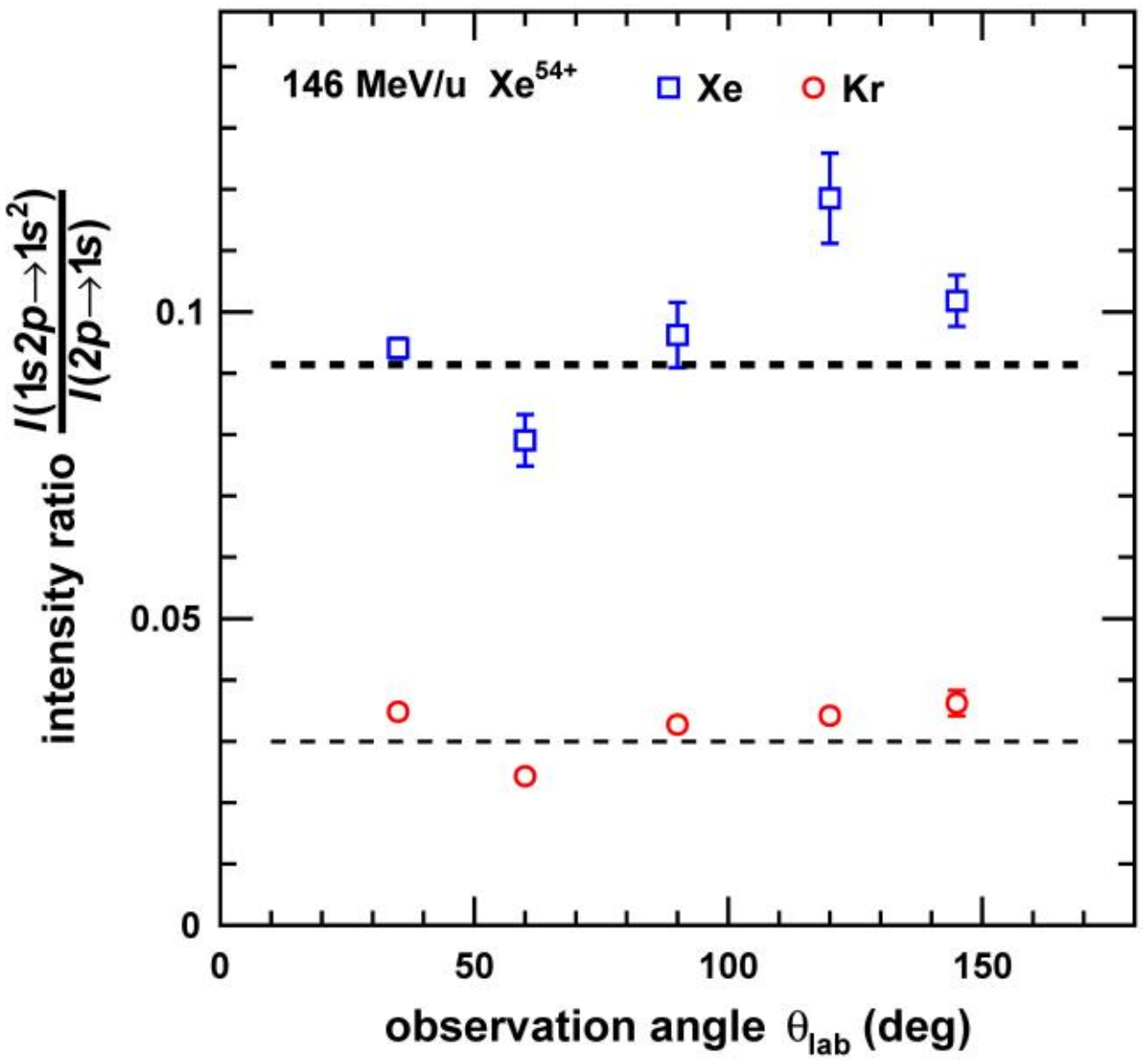

FIG. 10. Experimental intensity ratios $I(1s2p \rightarrow 1s^2)/I(2p \rightarrow 1s)$ of $Xe^{52+*}$ and $Xe^{53+*}$ ions as functions of the observation angle $\theta_{lab}$. Results are shown for collisions of 146 MeV/u $Xe^{54+}$ ions with Kr (red circles) and Xe (blue squares) targets. The error bars are estimated from the statistics of data. The weighted average values of the intensity ratios (i.e. the overall intensity ratios) are indicated by dashed lines.

Additionally, the population distribution of the excited *L*, *M*, and *N*-shell levels of projectile ions produced by the NRC and NRDC processes can be deduced from the intensity ratios of I[K$\alpha_1$(+M2)]/I[Ly-$\alpha_2$(+M1)], I(Ly-β)/I(Ly-α), I(Ly-γ)/I(Ly-α), and I(Kα)/I(Ly-α).

Figure 8 shows the result for emission pattern of the K$\alpha_1$(+M2) transition normalized to the Ly-$\alpha_2$(+M1) intensity. The angular distribution is obvious anisotropy. The anisotropy parameter $\beta_{20}$ and the total cross-section ratios $\sigma^{\text{total}}_{K\alpha_1(+M2)}/\sigma^{\text{total}}_{\text{Ly-}\alpha_2(+M1)}$ were extracted by fitting the experimental results from the formula of the angular distribution of the intensity ratio between K$\alpha_1$(+M2) and Ly-$\alpha_2$(+M1) transitions [17,35]. From this procedure, values of $\beta_{20}$ =0.41±0.04 and $\sigma^{\text{total}}_{K\alpha_1(+M2)}/\sigma^{\text{total}}_{\text{Ly-}\alpha_2(+M1)}$=0.03±0.01 were obtained. The quoted uncertainty is entirely of statistical origin. This finding implies that the double-electron capture is two orders of magnitude smaller than the single-electron capture. The determined anisotropy parameter is in turn related to the alignment of the corresponding magnetic sublevels whose exact form depends on the transition under consideration. As already mentioned for the NRDC of bare xenon, two unresolved levels of $(1s_{1/2}2p_{3/2})_1$ and $(1s_{1/2}2p_{3/2})_2$ are in principle populated and thus contribute to the K$\alpha_1$(+M2) transition. Therefore, there is impossible to deduce the alignment parameters and the population ratios of the magnetic sublevels for individual states by this fitting procedure. This is different to the Ly-$\alpha_1$ transition resulting from the NRC process because the alignment parameter for the $2p_{3/2}$ state is directly extracted from the angular distribution of the Ly-$\alpha_1$ line [17,24,25]. For comparison, the angular distribution of the Ly-$\alpha_1$ has large negative value of the anisotropy parameter which reflects the significant nonstatistical population of magnetic sublevels of the $2p_{3/2}$ state.

As can be seen in Fig. 3, different fine-structure components of the Ly-β and Ly-γ transitions were not distinguished in the present experiment. We analyzed angular dependences of the intensity ratios $I(\text{Ly-}\beta)/I(\text{Ly-}\alpha)$ and $I(\text{Ly-}\gamma)/I(\text{Ly-}\alpha)$. Figure 9 presents these ratios for 95 MeV/u $Xe^{54+}$→Kr collisions. Both ratios showed approximate isotropy across all observation angles, yielding weighted mean values for I(Ly-β)/I(Ly-α) and I(Ly-γ)/I(Ly-α) as shown in Fig. 9, respectively. The experimental findings indicate that the Ly-α, Ly-β, and Ly-γ transitions have the similar angular distribution. Our previous studies [24,25] observed that the Ly-$\alpha_1$ transition changes from strongly anisotropic to nearly isotropic as the projectile energy increased.

Moreover, the overall intensity ratios imply that the captured electrons preferentially populate lower states of $Xe^{53+*}$ resulting from the NRC process. To compare with the measured ratios, the theoretical results of relative cross sections $\sigma^{\text{REA}}_{M-\text{NRC}}/\sigma^{\text{REA}}_{L-\text{NRC}}$=0.27 and $\sigma^{\text{REA}}_{N-\text{NRC}}/\sigma^{\text{REA}}_{L-\text{NRC}}$=0.08 of NRC to the *L*-, *M*-, and *N*-shell of the projectile in 95 MeV/u $Xe^{54+}$→Kr collisions are calculated by the REA approach without considering the cascade feeding effect. Approximate agreement between the experimental and theoretical results can be seen.

The relative population of projectile excited states produced by NRDC can be deduced by the intensity ratios $I(1snp \to 1s^2)/I(np \to 1s)$ ($n$=2, 3, and 4) of $Xe^{52+*}$ and $Xe^{53+*}$ ions in collisions of $Xe^{54+}$ with Kr and Xe at projectile energies of 95-197 MeV/u. Figure 10 illustrates angular distributions of the intensity ratios $I(1s2p \to 1s^2)/I(2p \to 1s)$ in collisions of 146 MeV/u $Xe^{54+}$ ions with Kr and Xe targets as an example. The overall intensity ratios for these x-ray lines were determined due to nearly isotropic behavior of observed ratios. The uncertainties of the ratios were estimated from the statistics of data. As seen from the figure 10, the intensity ratio $I(1s2p \rightarrow 1s^2)/I(2p \rightarrow 1s)$ for Xe target is almost three times larger than that for Kr target. To compare with the experimental results, theoretical relative cross sections $\sigma_{KL}^{\text{IEA}}/\sigma_{L}^{\text{REA}}$ of the NRDC into the $KL$ shells and the NRC into the $L$ shell are 0.06 and 0.02 for Xe and Kr targets, which are calculated respectively using the IEA and the REA without considering the cascade feeding effect.

Combined with the results of $I(1s3p \rightarrow 1s^2)/I(3p \rightarrow 1s)$ and $I(1s4p \rightarrow 1s^2)/I(4p \rightarrow 1s)$ for both targets at other energies, the features of the energy and target dependence of the population of excited projectile levels produced by the NRDC were revealed [23]. The contribution of NRDC is absent at 197 MeV/u for both Kr and Xe targets. At the energy of 95 or 146 MeV/u, the three intensity ratios $I(1snp \rightarrow 1s^2)/I(np \rightarrow 1s)$ ($n$=2, 3, and 4) are nearly equal for Kr, the one $I(1s2p \rightarrow 1s^2)/I(2p \rightarrow 1s)$ is obvious larger that the other two ratios for Xe [23]. In addition, the $1s2p \rightarrow 1s^2$ transition has similar angular emission patterns to that of the $2p \rightarrow 1s$ line, as well as the lines of $1s3p \rightarrow 1s^2$ and $1s4p \rightarrow 1s^2$ [23].

It is noted that the obtained experimental results include the cascade feeding contribution of nonradiative electron capture into higher excited states. Nevertheless, the present theoretical results are given without considering the cascade feeding because (magnetic) sublevel cross sections are not well calculated yet. Furthermore, experimental studies on the NRC and NRDC in relativistic collisions of bare and H-like high-Z ions with heavy atoms are few up to now. The NRC and NRDC experimental results in the present work can test theoretical methods and may drive additional experimental studies in this field.

## 4 Summary and outlook

We studied the nonradiative single- and double-electron capture in collisions of 95-197 MeV/u bare xenon ions with krypton and xenon gaseous atoms at the HIRFL-CSR storage ring. The population distribution of excited states in H-like and He-like heavy ions produced by the NRC and NRDC processes have been determined by analyzing Doppler-shift x rays emitted from fast-moving down-charged projectiles.

The relative population of excited L, M, and N-shell levels in $Xe^{53+*}$ resulting from NRC were investigated by the nearly isotropy intensity ratios of $I(\text{Ly-}\beta)/I(\text{Ly-}\alpha)$ and $I(\text{Ly-}\gamma)/I(\text{Ly-}\alpha)$. Compared with the relative cross sections calculated by the REA approach, there are noticeable difference. Moreover, the target electrons are populated more into the lower excited levels.

The anisotropy parameters and alignment parameters for two unresolved $(1s_{1/2}2p_{3/2})_1$ and $(1s_{1/2}2p_{3/2})_2$ states of $Xe^{52+*}$ produced by the NRDC mechanism were determined by the angular distributions of the $K\alpha_1$(+M2) transition. The difference following NRC and NRDC into the $L$-shell levels of projectiles is found when compared to anisotropy parameters of the Ly-$\alpha_1$ transition of $Xe^{53+*}$ produced by NRC. In addition, the obtained total intensity ratios of $I(K\alpha_1 + M2)/I(\text{Ly-}\alpha_2 + M1)$ and $I(1s2p \rightarrow 1s^2)/I(2p \rightarrow 1s)$ indicate that the NRDC is significantly smaller than the NRC for all the collision cases considered above.

The Doppler-shift x-ray measurements presented here provide critical insights into NRC and NRDC dynamics in relativistic heavy-ion collisions. By resolving the angular dependence of characteristic x-ray emissions from $Xe^{53+}$ and $Xe^{52+}$ ions, we quantify state-selective population mechanisms and reveal strong Z-dependent effects in anisotropy parameters, offering stringent tests for relativistic atomic collision theories.

**ACKNOWLEDGMENTS**

We thank all participating members of the accelerator department for their operation of the Heavy Ion Research Facility at Lanzhou–Cooling Storage Ring. This work was supported by the National Key Research and Development Program of China under Grant No. 2022YFA1602500, the "Light of West China" Program of Chinese Academy of Sciences under Grant No. xbzglzb2022004, the National Natural Science Foundation of China under Grants No. 12375263, and Gansu Provincial Talent Young Program under Grant No. 2025QNGR66.